# Exploring motivation and teamwork in a large software engineering capstone course during the coronavirus pandemic

Yngve Lindsjørn, Steffen Almås, Viktoria Stray

Department of Informatics, University of Oslo, Norway {ynglin, steffa, stray}@ifi.uio.no

**Keywords:** Teamwork, Agile practices, Software Engineering, Covid-19, Capstone course

**ABSTRACT:** In the spring of 2020, the Department of Informatics covered a 20 ECTS capstone course in Software Engineering, mainly focusing on developing a complex application. The course used active learning methods, and 240 students were working in 42 cross-functional, agile teams. The pandemic caused by the coronavirus had a significant impact on the teaching given by the University of Oslo, as all physical education and collaboration among the teams had to be digital from March 12. At the end of the semester, we conducted a survey that focused on 1) aspects of teamwork (e.g., communication and coordination in the teams) and the relation to team performance (e.g., the application product) and 2) the students' motivation and ability to cooperate through digital platforms. A total of 151 respondents in 41 agile student teams answered the survey. This study aimed to investigate how the teamwork and motivation of the students were affected by having to work virtually. The results are compared to results from the same course in 2019 and a similar survey on 71 professional teams published in 2016. Our results show that the teamwork was evaluated similarly to both the evaluation of survey conducted in 2019 and on the professional teams in 2016. The motivation among the students remained high, even though they had to collaborate virtually.

## 1 INTRODUCTION

Working in teams in software development is essential (Chow & Cao, 2008). Developers work in teams because, "in software development specifically, the speed, frequency, complexity, and diversity of changes needed for modern software-rich systems mean that teams are essential." (Skelton & Pais, 2019). Agile software development is now the common practice and provides values and principles for producing working software rapidly while responding effectively to change.

The Department of Informatics at the University of Oslo offers a 20 ECTS capstone software engineering course where working in teams is a central part. Due to the coronavirus situation, the Campus was closed down on March 12th, 2020. This course's consequence was that the project work (which started ten days before the lockdown) and collaboration between the team members had to be carried out using digital tools.

Other research on the effect of the coronavirus situation in higher education in Norway includes some relevant articles from the 2020 NIKT conference (Norwegian conference for ICT-research and education). One of the findings in (Lorås et al., 2020) is that "informal learning spaces are essential to students yet challenging to transfer effectively to the online environment." Hjelsvold et al., 2020 reported that some educators found certain aspects of online teaching to be better than when teaching physically on Campus. When facing such sudden changes, the educators were good at collaborating and exchanging pedagogical experience.

This article aims to investigate the effect of not being able to physically meet when developing software in student teams. The research question is formulated as follows: *How did the coronavirus situation and the Campus shutdown affect the teamwork and the motivation of the students in a large software engineering capstone course?*

This study uses data from the course in spring 2020 (42 teams) and data from the same course in 2019 (39 teams) for comparison. Central in the study is a survey given to the students at the end of the semester. In 2019 and 2020, there were questions about teamwork using the teamwork quality (TWQ) concept and the relationship to team performance. TWQ was initially developed and used by Hoegl and Gemuenden (2001) and further applied by Lindsjørn et al. (2016) and Lindsjørn et al. (2018). The TWQ constructs measure the quality of interactions within a team and consist of six variables; *communication*, *coordination*, *balance of member contribution*, *mutual support*, *effort*, and *cohesion*. Team Performance consists of the variables *effectiveness* and *efficiency*, and the Team Members' Success consists of *work satisfaction* and *learning*. Effectiveness refers to the expectation regarding product quality, while efficiency refers to the expectations regarding project quality, such as time and cost (Lindsjørn et al., 2016).

Due to the coronavirus situation in 2020, we decided to include additional questions regarding the digital learning environment using the concept of sociability, which refers to the extent to which digital tools are perceived to help cope with the distributed teamwork (Kreijins et al., 2007). Some key attributes of the concept of sociability are trust within the team, belonging, and relationship.

## 2    THE COURSE

The 20 ECTS capstone software engineering course is mandatory for most students at the Department of Informatics and has approximately 250 students each semester. The course starts with eight weeks of intensive lectures (see Table 1) and group sessions. During these weeks, the students also prepare for teamwork, and teams are formed with ideally six students in each team (Løvold et al., 2020). The students submit a survey regarding their motivation, background, and up to two fellows they want as team members. Based on the survey response, the students are assigned to a team by the course lead.

|        |    | Lecture |
|--------|----|---------|
| Week 1 | 1  | Introduction to the course |
|        | 2  | The Basics of Android Studio and Kotlin |
| Week 2 | 3  | More on Android Studio |
|        | 4  | More on Kotlin |
| Week 3 | 5  | API, data formats, HTTP-requests and Proxy-servers |
|        | 6  | Teamwork, agile methodologies and project work |
| Week 4 | 7  | Agile practices |
|        | 8  | Basic Principles of Testing |
| Week 5 | 9  | Secure System Development |
|        | 10 | Modelling and object-oriented principles |
| Week 6 | 11 | Architecture and Technical Debt |
|        | 12 | From Theory to Practice – the project from A to Z |
| Week 7 | 13 | Application Programming Interface (API) |
|        | 14 | Development of Android apps and use of patterns |
| Week 8 | 15 | Universal Design |
|        | 16 | Evaluation Method / Research Methods |

*Table 1 – Lectures in the Software Engineering capstone course*

During the project period of 12 weeks, the student teams were assigned to make a mobile weather app on the Android platform using data from The Norwegian Meteorological Institute´s API (MET, 2020). There were six weather cases made in collaboration with the Norwegian Institute of Meteorology. The cases were titled as follows:
- Water movements in the oceans
- Forecasts of landslides and avalanches
- Air quality in municipalities
- Predictions of climate and climate change
- Drones and airspace
- Open case – use weather data and design your own case

The students were given introductions to agile methods and software engineering practices, and the teams were free to select development methodology and practices in their teamwork. Practically all the teams used their own adaptations of the agile process models Scrum and Kanban and applied agile practices such as sprints, daily meetings, sprint planning, and retrospective meetings. The majority of the teams had a designated Scrum master, and some teams rotated the Scrum master role during the period.

The students were free to select collaboration tools to work virtually, except GitHub, which was mandatory for Version Control.

An essential part of this course is the mandatory presentations held at the end of the project. During these presentations, the students share their thoughts and reflect on their experience of working in teams. They also presented their final product by demonstrating their application online, where everyone was invited. Several teaching assistants and lecturers were present during these presentations, making it possible for the students to share their opinions directly with the course lead. As a part of the grading, the students delivered a comprehensive report (together with the app) at the end of the project. In the reports, the team members elaborated on the process leading up to the final product and described both the product (technical and non-technical aspects) and the process. They also reflected on the coronavirus situation and how they were affected.

## 3  METHOD

In June 2020, a survey was conducted after the student teams had submitted and presented their project. The survey was published on June 1st, 2020, and was open for one month (the last response was received on July 1st). The survey consisted of a total of 91 items. We included 10 questions measuring sociability (digital learning environment) and 61 questions measuring TWQ and team performance. In addition, we included 20 items about the tools and background questions like gender and age, study program, and previous experience in agile development. A total of 151 students out of 240 responded, making the response rate 63%. In May and June 2019, when the same survey was conducted, the response rate was higher as the students answered the forms physically at the university during the mandatory presentations (the response rate in 2019 was 98%). Each item in the TWQ and team performance model, and items in the digital learning (Sociability model) of the questionnaire were statements. The respondents indicated their personal views for each statement on a Likert-scale from 1 (strongly disagree) to 5 (strongly agree).

*Cronbach alpha* is a statistic for internal-consistency reliability alpha values and should be higher than 0.7 to be satisfactory (Nunnaly and Bernstein (1994). All alpha values in all constructs were satisfactory, except the TWQ construct Balance of member Contribution (3 items). In a similar study conducted on professional teams, the same construct was the only construct with the alpha value below 0.7 as well (Lindsjørn et al. (2016).

## 4  RESULTS AND DISCUSSION

All teams found appropriate tools useful in teamwork. Zoom was used during the digital lectures and supervision, so naturally, this tool's usage among students was high. The most popular tool used was Slack (used by 81 %). The use of Slack has been found valuable in agile distributed teams because it increases team awareness (Stray, 2020). Next, the survey respondents reported using Zoom (74 %), Facebook Messenger (42 %), and Discord (37 %). Other tools were also used, such as Google Disk, Microsoft Teams, Trello, Monday, Notion, and Skype. Most of the teams used more than one tool. Some teams started chatting on Slack but realized that video meetings were more effective: «... *Since we could not meet physically and discuss at the University, we met digitally instead. Through trial and error, we eventually found a balance that worked for everyone. In the start, we mostly used Slack to discuss, but gradually we discovered meeting over video (Zoom) was more effective*." Some teams reported that they were not affected by the situation as stated in one of the reports: "*Generally, we feel that the project was not very affected by the coronavirus situation, and we managed to work well together with the help of digital tools and good communication*."

Table 2 shows that the results are similar for both 2019 and 2020. We see that the mean values of the TWQ variables are slightly higher in 2019 than in 2020, in particular, *Communication* with a difference of 0,18 in 2020 and Effort with a difference of 0,13. The mean values for the team performance variables, however, are higher in 2020 than in 2019, both Effectiveness (product quality) and Efficiency (project quality). Compared to the study conducted on professional teams (Lindsjørn et al., 2016), the mean values of the student teams were slightly higher (both in 2019 and 2020) than the values of the study of professional teams. The variance among the team members' evaluation of all variables was significantly higher in the students' teams than in the professional teams, with a standard deviation of 0,45 on average in the student teams and 0,30 on average among the professional teams.

All values in Table 2 are calculated on the team level: the aggregated values of all team members' evaluation, while the values in Table 3 are calculated on the individual level. This does not affect the mean values, only the standard deviation values. For the TWQ and team performance model, the values are presented on the variable level (the variable communication has, e.g., 10 items – See Table 2). In contrast, the only construct in the Sociability model (digital learning environment) is presented on the item (question) level (see Table 3).

| Construct | Variable | No. Items | 2019 Mean | 2019 Std. Dev. | 2020 Mean | 2020 Std. Dev | Difference Mean | Difference Std. Dev |
|---|---|---|---|---|---|---|---|---|
| Team Quality (TWQ) | Communication | 10 | 4,17 | 0,37 | 3,99 | 0,51 | -0,18 | 0,14 |
| | Coordination | 4 | 4,05 | 0,40 | 3,98 | 0,45 | -0,07 | 0,05 |
| | Mutual support | 7 | 4,42 | 0,37 | 4,33 | 0,47 | -0,09 | -0,10 |
| | Cohesion | 10 | 4,26 | 0,45 | 4,20 | 0,45 | -0,06 | 0,00 |
| | Effort | 4 | 3,86 | 0,65 | 3,73 | 0,62 | -0,13 | -0,03 |
| | Balance of contribution | 3 | 4,25 | 0,41 | 4,24 | 0,50 | -0,01 | 0,09 |
| Team members' success | Work satisfaction | 3 | 4,28 | 0,44 | 4,37 | 0,40 | 0,09 | -0,04 |
| | Learning | 5 | 4,41 | 0,47 | 4,42 | 0,45 | 0,01 | -0,02 |
| Team performance | Effectiveness | 10 | 3,86 | 0,42 | 4,03 | 0,34 | 0,17 | -0,08 |
| | Efficiency | 5 | 3,81 | 0,60 | 3,98 | 0,55 | 0,17 | -0,05 |

*Table 2 – Descriptive statistics of investigated variables*

| No. | Item | Mean | Standard deviation |
|---|---|---|---|
| 1 | The virtual learning environment enables me to easily contact my teammates | 3,96 | 1,07 |
| 2 | I do not feel lonely in the virtual learning environment | 3,62 | 1,18 |
| 3 | The virtual learning environment enables me to get a good impression of my teammates | 3,33 | 1,15 |
| 4 | The virtual learning environment allows spontaneous informal conversations | 3,57 | 1,24 |
| 5 | The virtual learning environment enables us to develop into a well performing team | 3,60 | 1,03 |
| 6 | The virtual learning environment enables me to develop good work relationships with my teammates | 3,44 | 1,15 |
| 7 | This virtual learning environment enables me to identify myself with the team | 3,51 | 1,11 |
| 8 | I feel comfortable in the virtual learning environment | 3,87 | 1,02 |
| 9 | The virtual learning environment allows for non-task-related conversations | 3,44 | 1,19 |
| 10 | The virtual learning environment enables me to make close friendships with my teammates | 2,78 | 1,28 |

*Table 3 – Descriptive statistics of virtual learning environment items*

Table 3 shows the mean and standard deviation values of the evaluation of the virtual learning environment items. All the 9 first items' mean values are between 3 and 4 (3,6 on average). The evaluation of item 10 (making close friendship) was significantly lower than the other items' evaluation. Similar findings were found in Kreijns et al., 2007.

It is remarkable how the student teams have adapted so quickly to virtual teamwork and that they evaluated both aspects of the teamwork, satisfaction of work, and product similar to the student teams in 2019 were they met physically. The results presented in Table 3 (virtual learning environment) also support the fact that the student teams could collaborate well regarding the project. However, it shows that it was hard to make close friendships with teammates. This indicates that working remotely has a more negative impact on social aspects than the teamwork among students in capstone courses like this.

Most of the teams reported during the presentations that the case they had chosen was engaging, which raised the motivation. All the teams were asked the same question after the presentation of the project: "*How did the coronavirus situation (closing down the Campus) and the fact that the grade was only passed/not passed influence your motivation in the course*"? The most common answer was: "*The motivation became lower at once, but when we really started to work (digitally) together as a team, we just wanted to make a good app, write a good report and learn to use some agile practices during the teamwork.*"

Most of the student reports reflected this fact, and some teams also reported advantages. Here is an example: *"... We discovered some advantages of having digital meetings, i.e., fewer excuses to skip, and using digital entertainment mediums such as "Jackbox" was a nice way to replace physical Game nights."* Another example: *"... We were able to collaborate well despite the social distance. The digital meetings went well and were probably more effective than what psychical meetings would have been. Since we met digitally, we delegated the tasks more clearly than we would if we met physically"*.

## 5    CONCLUSION

This article has addressed the following research question: How did the Corona situation and the Campus shutdown affect the teamwork and the motivation of the students in a large capstone course working in teams? Our findings show that the teamwork worked well despite working remotely. Though many students' motivation dropped just after the lockdown, the motivation increased when they started to work (digitally) together as a team. One of the reasons that the teamwork worked well and the motivation increased is that the students found good digital collaboration tools. Another reason was that they found the cases to be exciting and challenging.